\documentclass[aps,prl,twocolumn,showpacs,preprintnumbers,amsmath,amssymb,superscriptaddress]{revtex4-1}
\usepackage{color}
\usepackage{graphicx}% Include figure files
\usepackage{dcolumn}% Align table columns on decimal point
\usepackage{bm}% bold math

\usepackage[hypertex]{hyperref}
\newcommand{\nc}{\newcommand}
\nc{\be}{\begin{eqnarray}}
\nc{\ee}{\end{eqnarray}}
\nc{\bea}{\begin{eqnarray}}
\nc{\eea}{\end{eqnarray}}
\nc{\bean}{\begin{eqnarray*}}
\nc{\eean}{\end{eqnarray*}}
\nc{\mb}{\mbox}
\nc{\rnc}{\renewcommand} 
\nc{\vk}{\bf k}
\nc{\vx}{\mb{\bf x}}
\nc{\br}{\mb{\bf r}}
\nc{\bv}{\mb{\bf v}}
\nc{\bp}{\mb{\bf p}}
\nc{\ve}{\mb{\bf e}}
\nc{\vz}{\hat {\mb{\bf z}}}
\nc{\vp}{\mb{\boldmath$p$}}
\nc{\vb}{\mb{\boldmath$b$}}
\nc{\rr}{\mb{\boldmath$r$}}
\nc{\vR}{\mb{\boldmath$R$}}
\nc{\vj}{\mb{\boldmath$j$}}
\nc{\vg}{\mb{\boldmath$g$}}
\nc{\vm}{\mb{\boldmath$m$}}
\nc{\vd}{\mb{\boldmath$d$}}
\nc{\hd}{\mb{\boldmath$\hat{d}$}}
\nc{\vD}{\mb{\boldmath$D$}}
\nc{\vF}{\mb{\boldmath$F$}}
\nc{\vG}{\mb{\boldmath$G$}}
\nc{\vI}{\mb{\boldmath$I$}}
\nc{\vW}{\mb{\boldmath$W$}}
\nc{\x}{\mb{\boldmath$x$}}
\nc{\A}{\mb{\boldmath$A$}}
\nc{\va}{\mb{\boldmath$a$}}
\nc{\vv}{\mb{\boldmath$v$}}
\nc{\vq}{\mb{\boldmath$q$}}
\nc{\vn}{\mb{\boldmath$n$}}
\nc{\vJ}{\mb{\boldmath$J$}}
\nc{\vS}{\mb{\boldmath$S$}}
\nc{\vs}{\mb{\boldmath$\sigma$}}
\nc{\vE}{\mb{\boldmath$E$}}
\nc{\vB}{\mb{\boldmath$B$}}
\nc{\vM}{\mb{\boldmath$M$}}
\nc{\vL}{\mb{\boldmath$L$}}
\nc{\vpsi}{\mb{\boldmath$\psi$}}
\nc{\vphi}{\mb{\boldmath$\varphi$}}
\nc{\Vphi}{\mb{\boldmath$\phi$}}
\nc{\Vomega}{\mb{\boldmath$\Omega$}}
\nc{\ipsi}{\it{\Psi}}
\nc{\vepsilon}{\mb{\boldmath$\epsilon$}}
\nc{\valpha}{\mb{\boldmath$\alpha$}}
\nc{\vgamma}{\mb{\boldmath$\gamma$}}
\nc{\vomega}{\mb{\boldmath$\omega$}}
\nc{\vmu}{\mb{\boldmath$\mu$}}
\nc{\vt}{\mb{\boldmath$\tau$}}
\nc{\vT}{\mb{\boldmath$T$}}
\nc{\vpi}{\mb{\boldmath$\pi$}}
\nc{\nab}{\bm{\nabla}}
\nc{\ov}{\overline}
\nc{\cdott}{\!\cdot\!}
\nc{\cdottt}{\!\!\cdot\!}
\nc{\LL}{\Big{\langle}}
\nc{\RR}{\Big{\rangle}}
\nc{\LR}{\Bigm{|}}
\nc{\vP}{\mb{\boldmath$P$}}
\nc{\nnn}{\nonumber\\}
\rnc{\figurename}{FIG.}

\nc{\psibar}{\overline{\psi}}
\nc{\cbar}{\overline{c}}
\nc{\intx}{\int d^4x}
\nc{\inty}{\int d^4y}
\nc{\intk}{\int \frac{d^4k}{(2\pi)^4}}

\nc{\eq}[1]{(\ref{#1})}

\def\nomura#1{\textcolor{black}{#1}}

\begin{document}

\title{
Charge-Induced Spin Torque in Anomalous Hall Ferromagnets
}

\author{Kentaro Nomura}
\affiliation{
Institute for Materials Research, Tohoku University, Sendai 980-8577, Japan
}

\author{Daichi Kurebayashi}
\affiliation{
Institute for Materials Research, Tohoku University, Sendai 980-8577, Japan
}

\date{\today}

\begin{abstract}
We demonstrate that spin-orbit coupled electrons in a magnetically doped system exert a spin torque on the local magnetization, without a flowing current, when the chemical potential is modulated in a magnetic field. The spin torque is proportional to the anomalous Hall conductivity, and its effective field strength may overcome the Zeeman field.
Using this effect the direction of the local magnetization is switched by gate control in a thin film.  
This charge-induced spin torque is essentially an equilibrium effect in contrast to the conventional current-induced spin-orbit torque, and, thus, devices using this operating principle possibly have higher efficiency than the conventional one. In addition to a comprehensive phenomenological derivation, we present a physical understanding basing on a model of a Dirac-Weyl semimetal, possibly realized in a magnetically doped topological insulator. The effect might be realized also in \nomura{nanoscale} transition materials, complex oxide ferromagnets, and dilute magnetic semiconductors. 

\end{abstract}

\maketitle

%\noindent
%introduction
%\section{Introduction}

{\it Introduction.}---
The electric control of spin magnetization aims to be used in next-generation magnetic devices, allowing information to be written electronically.
%The electric control of spin-magnetization has been aimed for application to next generation magnetic devices where the information is written electrically. 
Spin-transfer torque random-access memory has emerged as a potential candidate for such versatile devices:
a spin-polarized current exerts a spin-transfer torque on the magnetization and switches the direction via the exchange interaction
% between itinerant electrons and local spins
\cite{STT-review}. 
It is known that the driving spin-polarized current needs to exceed a threshold current, and a noncollinear magnetization structure such as spin valves, tunnel junctions, or domain walls is required. 
These might be central issues for low-power-consumption magnetic-recording devices.
The spin-orbit torque has been recently proposed to control the magnetization direction without noncollinear configurations.
The threshold current density is $\sim10^6$ A/cm$^2$ for a number of magnetic materials\cite{SOT-review}, 
so much effort has been made to search for materials having high efficiency
\cite{Pareek2007,Manchon2009,Miron2010}.

In this Letter, we shall propose an alternative mechanism to switch the magnetization by electrical means in anomalous Hall ferromagnets consisting of local spins and itinerant band electrons. 
The anomalous Hall effect (AHE) occurs in solids with broken time-reversal symmetry, typically in a ferromagnetic phase, as a consequence of spin-orbit coupling \cite{AHE-review}.
In particular, the intrinsic AHE originates to the spin-orbit coupled band structure and can be described in terms of Berry curvatures \cite{Berry-review}. 
In many cases, the intrinsic effect appears to be the dominant contribution to the AHE in the low-temperature, clean limit of metallic ferromagnets\cite{AHE-review}.
%This gives an intrinsic contribution to the Hall conductivity in the low-temperature, clean limit of metallic ferromagnets, and in many cases appears to be the dominant contribution to the AHE\cite{AHE-review}. 

We derive a generic expression of the spin torque term induced by the chemical potential modulation and a magnetic field in anomalous Hall ferromagnets, based on a comprehensive phenomenological argument. This torque is proportional to the anomalous Hall conductivity. When this torque effect overcomes the Zeeman effect, the magnetization can be controlled locally  pointing parallel and antiparallel to the external magnetic field, depending on the sign of the chemical potential modulation.
%develop the response theory of intrinsic anomalous Hall ferromagnets, which reveals coupling of the charge density of itinerant electrons and the local spin-magnetization in a magnetic field. By applying it to thin-film systems, a spin torque is induced by the gate control. The torque remains finite in the absence of electric currents, in contrast to the current induced spin-transfer torque. 
%
Devices using this operating principle are free from joule heating and, thus, possibly have a much higher efficiency than conventional ones.
As an example of spin-orbit coupled band electrons we consider a Weyl semimetal\cite{Wan2011a,Burkov2011,Weyl-review,Weyl-review2,Dirac-review} realized in magnetically doped topological insulator materials\cite{Kurebayashi2014}, where a physical understanding and an estimation of the effective field are given.
%We demonstrate and examine the effect in a lattice model simulation.

%In this Letter we shall derive a spin torque term from a comprehensive phenomenological argument for a ferromagnet which consists of spin-orbit coupled itinerant electrons and local spins.

%\section{Charge density and Magnetization coupling in Anomalous Hall ferromagnets}
{\it Charge-induced spin torque}.---
In isotropic ferromagnets, the off-diagonal conductivity tensor $\sigma_{ij}$ ($i\neq j$)
may be expressed in the form\cite{AHE-review}
\bea
\sigma_{ij}^{}=\epsilon_{ijk}^{}\sigma_{\rm AHE}^{}\hat{M}_k^{}
\label{anomalous-hall}
\eea
in a vanishing magnetic field, 
where $\hat{\bm M}$ is the normalized directional vector of the magnetization and $\sigma_{\rm AHE}$ is the magnitude of the anomalous Hall conductivity. In the following we ignore the disorder effects and consider only the intrinsic contribution at zero temperature.

According to the Str\v{e}da formula\cite{Berry-review,Streda1982} the intrinsic Hall conductivity controls the charge density $n$ induced when a uniform magnetic field $\vB$ is applied: 
\bea
\sigma_{ij}^{}=-ec\epsilon_{ijk}^{}\frac{\partial n}{\partial B_k^{}},
\label{Streda-formula}
\eea
where
%$n$ is number of electrons, $\bm B$ is an external magnetic field, 
 $c$ is speed of light, and $e$ is the electron charge.
%Originally Str\v{e}da formula is a global relation, however, as far as the external magnetic field changes adiabatically, the local density of electrons satisfies the formula.
%
%
Combining these two relations, we obtain the relation between the electron density and the magnetization,
\bea
n_{\rm ind}=-\frac{\sigma_{\rm AHE}}{ec} \hat{\bm M}\cdot\bm B.
\label{n_ind}
\eea
In a magnetic field, the right-hand side of Eq.~(\ref{anomalous-hall}) is replaced by the $\sigma_{\rm AHE}\hat{M}_k+\sigma_{\rm OHE}\hat{B_k}$. However, the second term, the ordinary contribution $\sigma_{\rm OHE}\propto |{\bm B}|$, appears to be of second order in the field in Eq.~(\ref{n_ind}), and is ignored, since we focus on the linear response regime.
%Here we assume that the magnetization is hard enough and does not change its structure or amplitude by external magnetic fields.
%Equation~(\ref{n_ind}) indicates that charge is induced by the external magnetic field in a ferromagnet with the finite intrinsic anomalous Hall conductivity similarly to the quantum Hall systems\cite{Berry-review}.
% the time-reversal symmetry broken systems with finite value of the Berry curvature.
%Physically those systems are realized in ferromagnetic semiconductors and Weyl semimetals.
%Weyl semimetals are three-dimensional topological materials possessing magnetic monopoles in the momentum space.
%We are going to discuss on Weyl semimetals later as an example.
%According to the thermodynamical consideration which a conjugative variable for the number of electrons are the chemical potential, thermodynamical free energy is given by $-\int d^{3}x\,  n_{\rm ind}\delta\mu_F$.
These relations are derived in uniform systems, but when the magnetic field and magnetization vary slowly in space and time it is natural to assume that these are locally applicable. In the following $\sigma_{\rm AHE}$ denotes the magnitude of the anomalous Hall conductivity in the ideal uniform case.

In thermodynamics, the number of particles is conjugate to the chemical potential, described by the thermodynamic potential: 
 $-\int d^{3}x\,  n_{\rm ind}\delta\mu_F$.
By substituting Eq.~(\ref{n_ind}) into this relation, we derive a generic thermodynamic potential for charge and spin coupling in anomalous Hall ferromagnets as
\bea
\Omega_{\rm CS}=\int d^{3}x \frac{\sigma_{\rm AHE}}{ec}\delta \mu_F\hat{\bm M}\cdot\bm B,
\label{free_energy}
\eea
where $\delta \mu_F$ is a local chemical potential which is defined as a deviation from the Fermi energy.
When the magnetization is uniform while the chemical potential varies in space, Eq.~(\ref{free_energy}) can be rewritten as 
$\Omega_{\rm CS}=-\frac{1}{c}\int d^{3}x \bm A\cdot \bm j_{\rm AHE}$
where $\bm j_{\rm AHE}=\sigma_{\rm AHE}\bm E\times \hat{\bm M}$ is the anomalous Hall current, $\bm E=\bm \nabla \delta \mu_F/e$, and ${\bm A}$ is the vector potential.
Therefore the Eq.~(\ref{free_energy}) describes the anomalous Hall effect.
From a microscopic model, Eq.~(\ref{free_energy}) may be obtained by integration over the fermionic degrees of freedom.

%\section{Electrical manipulation of the magnetization}
%{\it Electrically induced spin torque.}---
Coupling between the modulation of the chemical potential and the magnetization direction described by Eq.~(\ref{free_energy}) indicates the possibility of the magnetization switching by gate tuning in a ferromagnetic thin-film.
%
%Equation (\ref{free_energy}) describes coupling between the chemical potential and magnetization if anomalous Hall conductivity is non zero,  $\sigma_{\rm AHE}\ne0$. This suggests the possibility of magnetization manipulation with the chemical potential tuning.
%We will discuss an electrical manipulation of magnetization based on Eq (\ref{free_energy}).
%Let us start with considering the total energy of local magnetic moments.
The coupling energy of the magnetization and an applied magnetic field is 
 $E_{\rm field}=E_{\rm Zeeman}+\Omega_{\rm CS}$.
%We have discussed from electron side of view, however Eq (\ref{free_energy}) can be also seen as the energy functional of local magnetic moments.
%Since we apply magnetic field, we also have Zeeman energy 
Here
$
E_{\rm Zeeman}=-\int d^{3}x  \rho_{s}^{} g_{\rm }^{}\mu_{B} S\hat{\bm M} \cdot \bm B,
$
is the Zeeman term,  and $\rho_{s}$, $g$, and $S$ being the density, the Lande factor, and spin of the magnetic moments, respectively.
%The total energy for magnetization coupling with a magnetic field is $E_{\rm field}=E_{\rm Zeeman}+\Omega_{\rm AHE}$.
%\bea
%\nonumber E_{\rm field}&=&E_{\rm Zeeman}+\Omega_{\rm AHE}\\
%&=&-\int d^{3}x\left[\hbar\gamma S\rho_{\rm M}^{}-\frac{\sigma_{\rm AHE}}{|e|c}\delta \mu_F(\bm x)\right]\bm B\cdot \hat{\bm M}
%\nonumber &=&-\int d^{3}x\left[\frac{\hbar\gamma}{v_{\rm unit}}\bm B\cdot\hat{\bm M}-\frac{\sigma_{\rm AHE}}{|e|c}\delta \mu_F(\bm x)\bm B\cdot \hat{\bm M}\right]\\
%
%&=&-\int \frac{d^{3}x}{v_{\rm unit}}\hbar \gamma\left(1-\frac{v_{\rm unit}}{\gamma\hbar}\frac{\sigma_{\rm AHE}}{|e|c}\delta\mu_F(\bm x)\right)\bm B\cdot \hat{\bm M}
%&=&-\int \frac{d^{3}x}{v_{\rm unit}}\hbar \gamma\left(1-\frac{\gamma_{\rm eff}(\bm x)}{\gamma}\right)\bm B\cdot \hat{\bm M}
%\eea
In addition to the Zeeman torque ${\bm T}_{\rm Zeeman}=g\mu_{B} S \bm B\times \hat{\bm M}$, there exists an induced torque term given by
\bea
 {\bm T}_{\rm CS}&=&- \frac{\delta \Omega_{\rm CS}}{\delta(\rho_{s}\hat{\bm M})}\times \hat{\bm M}
 \nonumber \\ &=&
 %=
 -\frac{\sigma_{\rm AHE}}{ec\rho_{s}}\delta \mu_F(\bm x)
 \bm B\times \hat{\bm M}
\label{torque}
\eea
in a magnetic field. This is the main finding of the present work.

In a thin film the, chemical potential can be locally tuned by gating. When $\sigma_{\rm AHE}$ and $\delta\mu_F$ are large enough, depending on the sign of $\delta\mu_F$, the local magnetization points parallel or antiparallel to the external magnetic field.
This torque is contrasted to the current-induced spin-transfer torque\cite{STT-review}, which can be expressed as ${\bm T}_{\rm STT}=\frac{\hbar S}{e}({\bm j}\cdot\nab)\hat{\bm M}$ in the adiabatic limit. For ${\bm T}_{\rm STT}$ a constant electric current ${\bm j}$ is needed which generates the joule heating, while ${\bm T}_{\rm CS}$ requires only the chemical potential modulation;
the effect is essentially dissipationless.
% which is essentially dissipationless. This is an advantage of the proposed operation.

%This torque should be contrasted to the current-induced spin transfer torque\cite{STT-review}, which can be expressed as ${\bm T}_{\rm STT}=\frac{\hbar S}{e}({\bm j}\cdot\nab)\hat{\bm M}$ in the adiabatic limit, and also to the spin-orbit torque.

%For ${\bm T}_{\rm STT}$ a constant electric current ${\bm j}$ is needed which generates the Joule heating, while ${\bm T}_{\rm CS}$ requires only the chemical potential modulation which is essentially dissipationless. This is an advantage of the proposed operation.
%${\bm T}_{\rm CS}$ can be introduced in the equilibrium states, and thus differs qualitatively also from spin-orbit torque\cite{Manchon2009} in which the non-equilibrium spin density, induced by a current and spin-orbit coupling, exerts the torque on the local magnetization.

%\begin{figure}[htbp]
%\includegraphics[width=0.8\linewidth]{fig-07.eps}
%\caption{Schematic image of the electrical magnetization manipulation.}
%\label{device}
%\end{figure}

%\section{Weyl semimetals}

{\it Microscopic derivation in Weyl semimetals}.---
In the rest of the Letter we consider the torque term Eq.~(\ref{torque}) from a microscopic point of view, where the system consists of spin-orbit coupled itinerant electrons and local spins, interacting via exchange coupling
\bea
 {\cal H}_{\rm exc}&=& JSx_{s}\hat{\bm M}(\bm x,t)\cdot \vs.
\label{exchange_interaction}
\eea
Here $J$ is the exchange coupling constant, $x_{s}=\rho_s a^3$ is the ratio of the magnetic dopants, $a^3$ being the volume of the unit cell, and $\vs=(\sigma_x,\sigma_y,\sigma_z)$ are Pauli matrices describing the electron spin degrees of freedom.
The torque induced by exchange coupling is given by $-JSa^3\langle\vs\rangle\times\hat{\bm M}$.
It has been proposed that in the presence of spin-orbit coupling, a flowing current produces a nonequilibrium spin density $\langle \vs \rangle_{\rm neq}$ and, thus, the spin-orbit torque: ${\bm T}_{\rm SOT}=-JSa^3\langle\vs\rangle_{\rm neq}\times\hat{\bm M}$ \cite{SOT-review,Pareek2007,Manchon2009}.
By contrast, Eq.~(\ref{torque}) is induced by the modulation of the chemical potential and an external magnetic field, where the spin density $\langle\vs\rangle$ is finite in equilibrium.

 As a concrete example of spin-orbit coupled ferromagnets we consider a Weyl semimetal, and show that a finite spin density is generated by chemical potential tuning.
% estimate the spin torque Eq.~(\ref{torque}).
A simplified model consists of a Dirac semimetal (DSM) and local spins of magnetic dopants\cite{Kurebayashi2014}.
This can be related to magnetically doped topological insulators\cite{TI-review} such as 
chromium-doped $\rm Bi_2Se_3$\cite{Yu2010,ex_BiSe-mag,JAP2012} or
chromium-doped $\rm (Bi,Sb)_2Te_3$\cite{J.Zhang2013}, where by doping Cr the strength of spin-orbit coupling is reduced and the original band gap may collapse at a certain range of the doping  ratio\cite{Kurebayashi2014,note3}.
The low-energy effective Hamiltonian is given as ${\cal H}_{\rm WSM}={\cal H}_{\rm DSM}+{\cal H}_{\rm exc}$. Here
\bea
 {\cal H}_{\rm DSM}&=&v_F\tau_z\vs\cdot\Big(-i\hbar\nab+\frac{e}{c}\A(\bm x,t)\Big)+e\phi(\bm x,t)
\eea
describes massless Dirac fermions in three dimensions\cite{TI-review,BiSe-model,BiSe-model2,Weyl-exp1,Weyl-exp2}, 
where $({\bm A},\phi)$ is the electromagnetic potential, $v_F$ is the velocity, and the chirality $\tau_z=\pm$ labels the two-degenerate Weyl nodes.
We note that this model differs from the Weyl semimetal phase proposed in a topological insulator multilayer\cite{Burkov2011}. 
In the latter system, the magnetization needs to point perpendicular to the layers, and, thus, the off-diagonal conductivity tensor cannot be expressed in the form of Eq.~(\ref{anomalous-hall}).

\begin{figure}[b]
\includegraphics[width=1\linewidth]{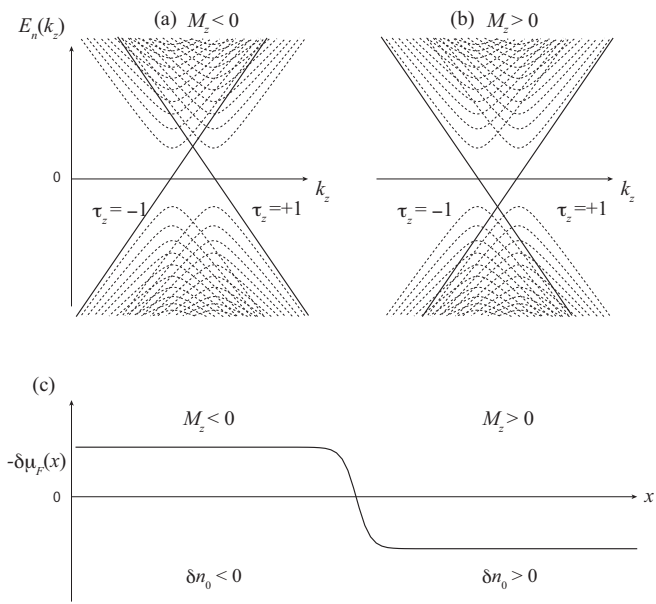}
\caption{
The energy dispersion of the Landau levels as a function of $k_z$ for (a) $\hat{M}_z<0$ and (b) $\hat{M}_z>0$.
(c) The modulation of the chemical potential as a function of the position.
}
\label{zero_mode}
\end{figure}

The proposed phenomena can be understood from the energetic point of view.
Here we consider Dirac fermions in a uniform magnetic field pointing $+z$ direction.
Defining the ladder operator $a=\sqrt{c/2\hbar eB_z}(\pi_x-i\pi_y)$ satisfying $[a,a^{\dag}]=1$, where ${\bm \pi}=-i\hbar\nab+(e/c){\bm A}+\tau_z (JSx_{s}/v_F)\hat{\bm M}$, the Haimltonian for a Weyl semimetal can be written as\cite{Weyl-review,Yang2011,Burkov2012}
\bea
 {\cal H}_{\rm WSM}
 =
 \tau_z \hbar v_F
 \begin{pmatrix}
% k_z+\tau_z K_0 & \sqrt{\frac{2eB_z}{\hbar c}}a\\
% \sqrt{\frac{2eB_z}{\hbar c}}a^{\dag} & -k_z-\tau_z K_0
  k_z+\tau_z\nomura{ \frac{JSx_{s}}{\hbar v_F}}\hat{M}_z & \sqrt{\frac{2eB_z}{\hbar c}}a\\
 \sqrt{\frac{2eB_z}{\hbar c}}a^{\dag} & -k_z-\tau_z \frac{JSx_{s}}{\hbar v_F}\hat{M}_z
 \end{pmatrix}.
 \nonumber \\
\eea
The zeroth Landau level states are obtained as $(0,|0\rangle)^t$ in the spinor representation, where $|n\rangle$ are the eigenstates of the number operator $a^{\dag}a$.
The energy dispersion is given by
\bea
 E_0(k_z)=-\tau_z\hbar v_F k_z\nomura{-J}Sx_{s}\hat{M}_z.
 \label{E_0}
\eea
Typical situations are illustrated in Fig.~\ref{zero_mode} (a) $\hat{M}_z<0$ (antiparallel to $\vB$) and 1 (b) $\hat{M}_z>0$ (parallel to $\vB$).
As represented by solid lines, the energies of the zeroth Landau level depend on the sign of $\hat{M}_z$.
By contrast, nonzero Landau levels
\bea
 E_n(k_z)=\pm\hbar v_F\sqrt{\Big(k_z+\frac{JSx_s}{\hbar v_F}\tau_z\hat{M}_z\Big)^2+\frac{2eB_z}{\hbar c}|n|},\ \ 
\eea
represented by the dashed lines in Fig. \ref{zero_mode},
are particle-hole symmetric, and the spectra do not differ for the opposite sign of  $\hat{M}_z$.
%In this way, we can interpret $\delta\mu_F$ as the effective chemical potential modulation for the zeroth Landau level states, induced by the relative directions of $\hat{\bm M}$ and $\vB$. 

When the modulation of the chemical potential $\delta\mu_F$ is introduced, electrons move from the high-potential region [$\delta\mu_F/(-e)>0$] to the low-potential region [$\delta\mu_F/(-e)<0$]. In the absence of the exchange interaction, all the Landau levels shift equally in energy by $\delta\mu_F$. In the presence of the exchange interaction, on the other hand, the density of electrons depends on the value of $\hat{M}_z$ as expected from Eq. (\ref{E_0}), and, thus, there is a correlation between $\hat{M}_z$ and $\delta\mu_F$.
To see this quantitatively, we count the number of electrons changed by $\hat{M}_z$ from the case of $\hat{M}_z=0$, fixing the magnetic field $B_z$ $(>0)$. Only the zeroth Landau level depends on the sign of $\hat{M}_z$, and, thus, modifies the density of electrons given by
\bea
 n_{\rm ind}&=& 
\frac{eB_z}{hc}\rho^{(1{\rm D})}_F\, \Delta E,
\label{n_ind_weyl}
\eea
where $\Delta E=JSx_{s}\hat{M}_z$ is the energy shift of the zeroth Landau level. Here $eB_z/hc$ is the degeneracy of the zeroth Landau level per area and $\rho^{(1{\rm D})}_F=2/2\pi\hbar v_F$ is the density of states of one-dimensional fermions, the product of them being the density of states in three dimensions.
Equation (\ref{n_ind_weyl}) is consistent with Eq. (\ref{n_ind}) 
and the anomalous Hall conductivity 
\bea
\sigma_{\rm AHE}=-\frac{e^{2}JSx_{s}}{2\pi^{2}\hbar^{2}v_F},
\label{weyl_ahe}
\eea
obtained from ${\cal H}_{\rm WSM}$\cite{Kurebayashi2014}.
To minimize the total energy, $\hat{\bm M}$ points in the direction of $-\vB$ in the high-potential region [$\delta\mu_F/(-e)>0$], while $\hat{\bm M}$ points in the direction of $+\vB$ in the low-potential region [$\delta\mu_F/(-e)<0$] as depicted in Fig. \ref{zero_mode} (c). The energy gained corresponds to $\Omega_{\rm CS}$.

The thermodynamic potential $\Omega_{\rm CS}$ can be also derived from microscopic field theory.
The effective action for the electromagnetic response in a Weyl semimetal has been derived as\cite{Weyl-review,Burkov2012,Burkov2013,Goswami2013}
%\bea
$
S_\theta=\frac{e^{2}}{4\pi^{2}\hbar c}\int dtd^{3}x\,\theta(\bm x,t)
\vE({\bm x},t)\cdot\vB({\bm x},t).
$
%\epsilon^{\mu\nu\rho\lambda} F_{\mu\nu} F_{\rho\lambda}.
%\label{theta_term}
%\eea
Here $\theta({\bm x},t)$ is the axion field
which is related to the magnetization direction in our model as
% in which the following condition is imposed in our model;
\bea
 \frac{1}{2}\nab\theta({\bm x},t)=\frac{JSx_{s}}{\hbar v_F}\hat{\bm M}({\bm x},t)=\frac{1}{\hbar v_F}\frac{\partial{\cal H}_{\rm exc}}{\partial\vs}.
\label{d-theta} 
\eea
The procedure to obtain the axion term $S_{\theta}$ is as follows.
First, in the Lagrangian formalism;
\bea
S_{\rm WSM}=\int dtd^3x\,\psi^{\dag}
\Big[i\hbar\partial_t-({\cal H}_{\rm DSM}+{\cal H}_{\rm exc})\Big]
\psi
,
\eea
we remove the exchange term ${\cal H}_{\rm exc}$ by the chiral gauge transformation;
$\psi\rightarrow e^{i\tau_z\theta/2}\psi$ where $\theta$ satisfies the condition Eq.~(\ref{d-theta}).
In the Grassman functional theory, the Jacobian $J_{\theta}$ is introduced by this transformation. After proper regularization\cite{Fujikawa}
the axion term is given by $S_{\theta}=S_{\rm WSM}-S_{\rm DSM}=-i{\rm ln}J_{\theta}$,
where $S_{\rm DSM}$ is $S_{\rm WSM}$ at $J=0$.
%We note that in Eq.~(\ref{d-theta}) $\hat{\bm M}$ does not have to be uniform but may depend on position and time.
The charge current is derived from $S_{\theta}$ as $\bm j=c\frac{\delta S_\theta}{\delta \bm A}=
%\bm j=
\frac{e^{2}}{4\pi^{2}\hbar}\bm \nabla\theta\times\bm E
%=\frac{e^{2JS}}{2\pi^{2}\hbar^{2}v_F}\bm M\times \bm E.
$. With Eq.~(\ref{d-theta}) we obtain the anomalous Hall conductivity Eq.~(\ref{weyl_ahe}).
%
%Namely the anomalous Hall conductivity in magnetically doped Weyl semimetals are obtained as 
%\bea
%\sigma_{\rm AHE}=\frac{e^{2}JSx_{\rm s}}{2\pi^{2}\hbar^{2}v_F}.
%\label{weyl_ahe}
%\eea
By substituting Eq.~(\ref{weyl_ahe}) for Eq.~(\ref{free_energy}), the thermodynamical potential $\Omega_{\rm CS}$ for Weyl semimetals can be obtained,
%Therefore the thermodynamical free energy Eq (\ref{free_energy}) for Weyl semimetals is obtained as
%\bea
%U_A=\int d^{3}x\frac{eJ}{2\pi^{2}\hbar^{2}v_F}\delta \mu_F\bm B\cdot \bm M.
%\eea
which is in agreement with $S_{\theta}$ when the electric field is written as  $\vE=\nab \delta \mu_F/e$.

With typical material parameters for $\rm Cr$-$\rm doped\ Bi_2Se_3$, we quantitatively estimate the ratio of the effective magnetic field to the external magnetic field ($x_s=0.1$, $J=2.0$eV, $\hbar v_F=2.2$eV\AA $^{-1}$ and $\rho_{s}=1.1\times 10^{-4}$\AA$^{-3}$)\cite{Yu2010,ex_BiSe-mag,BiSe-model,BiSe-model2,J.Zhang2013}
\bea
{\bm B}_{\rm eff}&\equiv& \frac{1}{g\mu_{\rm B}S\rho_{s}}\frac{\delta}{\delta \hat{\bm M}}(E_{\rm Zeeman}+\Omega_{\rm CS})
\nnn
&\approx&\Big[1-6.5\times\left(\frac{\delta \mu_F}{[\rm eV]}\right)\Big]\vB.
\label{effective_field}
\eea
If the chemical potential can be shifted by $\sim$0.2 eV, 
%the effective gyromagnetic ratio will be 1.6 times larger than the electron gyromagnetic ratio, and consequently 
the direction of the effective field $\vB_{\rm eff}$ and, thus, the magnetization are reversed.
In the above arguments we have neglected the Zeeman interaction for band electrons. 
In strongly spin-orbit coupled systems, it is known that the Lande factor of itinerant electrons can be larger than that in a vacuum.
Nevertheless, the typical energy scale of the Zeeman effect in $\rm (Bi,Sb)_2Te_3$ is $g^*\mu_B\approx 1$meV/T\cite{BiSe-model,BiSe-model2} which is negligibly small compered to the exchange interaction $JSx_s\approx 500$ meV in Eq.~(\ref{exchange_interaction}). \cite{Yu2010}

In the above argument we consider only the intrinsic contribution.
In the presence of disorder there exists the extrinsic contributions known as the skew-scattering and side-jump effects.
Generally, the Hall conductivity 
%is decomposed to 
consists of two parts $\sigma^{\rm I}_{ij}
=(\hbar e^2/4\pi){\rm Tr}[v_iG^+v_j(G^+-G^-)-v_jG^-v_i(G^+-G^-)]
$ and $\sigma^{\rm I\!I}_{ij}=(ie^2/4\pi){\rm Tr}[(x_iv_j-x_jv_i)(G^+-G^-)]$\cite{Streda1982},
where  $G^{\pm}=(E_F\pm i0-{\cal H})^{-1}$ and $E_F$ is the Fermi energy.
$\sigma^{\rm I}_{ij}$ is associated with states on the Fermi
surface.
$\sigma^{\rm I\!I}_{ij}$, on the other hand,
is the contribution of all states below the Fermi energy
and is a thermodynamic equilibrium property of the ferromagnet. In most cases, $\sigma^{\rm I}_{ij}$ corresponds to the extrinsic contribution\cite{AHE-review}, while $\sigma^{\rm I\!I}_{xy}$ the intrinsic contribution. The left-hand side of Eq.~(\ref{Streda-formula}) is $\sigma^{\rm I\!I}_{xy}$. 
The anomalous Hall conductivities, $\sigma_{ij}^{\rm I}$ and $\sigma^{\rm I\!I}_{xy}$, of a disordered Weyl metal have been computed in the model of topological insulator multilayers in Ref.~\cite{Burkov2014}. 
It was found that the extrinsic contribution to the anomalous Hall effect is absent as long as the Fermi level is sufficiently close to zero (Weyl nodes). This indicates that the charge-induced spin torque ${\bm T}_{\rm CS}$ of a Weyl semimetal is robust against disorder.

The mechanism of the induced effective field Eq.~(\ref{effective_field}) differs qualitatively from that proposed in a ferromagnet deposited on a topological insulator\cite{Garate2010}.
On the surface of a topological insulator, where Dirac-Weyl fermions demonstrate the quantum anomalous Hall effect\cite{TI-review}, a dissipationless Hall current $\vj$ produces a spin density $\langle\vs\rangle=-\frac{1}{ev_F}\vz\times\vj$ due to spin-momentum locking\cite{TI-review}, which gives an effective field and a torque\cite{Garate2010,Nomura2010,Tserkovnyak2015} as in the case of spin-orbit torque ${\bm T}_{\rm SOT}$\cite{Yokoyama2010,Ferreira,Nature2014,Shiomi2014,NatureMat2014}.
The effective field ${\bm B}_{\rm eff}$ generated by a current is pointing the in-plane direction, while the easy axis of the local magnetization is perpendicular to the surface.
To switch the magnetization, therefore, the current needs to exceed the threshold current, which might be challenging because a large current destroys the quantum Hall regime at the surface \cite{Garate2010}.

{\it Conclusion}.---
In this Letter, we derived a generic thermodynamic potential which describes coupling of the local spin magnetization and the charge density of itinerant band electrons in the three-dimensional anomalous Hall ferromagnets.
This indicates that a spin torque is locally induced by gate control without a flowing current. 
%Our argument is based on Eq.~(\ref{n_ind}) which is generic for intrinsic anomalous Hall ferromagnets, and the assumption that Eq.~(\ref{n_ind}) works not only for the uniform and static limit but also when physical quantities changes smoothly in space.
%The transient process is described, in the adiabatic limit, by the time derivative of Eq.~(\ref{n_ind}), $\dot{n}_{\rm ind}=-(\sigma_{\rm AHE}/ec)\vB\cdot \dot{\hat{\bm M}}$, where the magnetic field is assumed to be static and uniform. When the local magnetizations forms a texture, e.g. domain wall, vortex, or hedgehog, as described by $\hat{\bm M}({\bm x}-{\bm X}(t))$, the corrective motion of the texture is connected to the charge current ${\bm j}_{\rm ind}=(\sigma_{\rm AHE}/c)(\hat{\bm M}\cdot\vB)\dot{\bm X}(t)$ as total charge is conserved.
%
As an example, a Weyl semimetal was analyzed, and the strength of the effect was estimated. The torque term Eq.~(\ref{torque}) overcomes the Zeeman effect when the shift of the chemical potential is large enough, and, thus, the direction of the magnetization can be controlled by gate tuning. 
%The anomalous Hall conductivity of a Weyl semimetal is estimated as $\sigma_{\rm AHE}\approx 3.3\times 10^2[\Omega^{-1}{\rm cm}^{-1}]$.\cite{Kurebayashi2014}
The spin torque Eq.~(\ref{torque}) can be generated in ordinary ferromagnets with large anomalous Hall conductivity such as transition materials, complex oxide ferromagnets, and magnetic semiconductors.
\nomura{In practice, the region of switched magnetizations should be smaller than the Thomas-Fermi screening length as experimentally feasible and required for nanoscale devices.
}
%To generate the spin torque term Eq.~(\ref{torque}) a ferromagnet with large anomalous Hall conductivity, such as Iron, Cobalt, and Nickel ($\sigma_{\rm AHE}\approx 10^3 [\Omega^{-1}{\rm cm}^{-1}]$) might be suitable. 
The proposed mechanism of the induced spin torque in this work potentially has great advantage in application to low-energy-consumption nonvolatile memory devices.

\section{Acknowledgments}

This work was supported by Grant-in-Aid for Scientific
Research (No. 15H05854, No. 26107505 and No. 26400308) from the
Ministry of Education, Culture, Sports, Science and Technology (MEXT),
Japan.

\bibliography{3_Weyl_DW}

\begin{thebibliography}{10}

\bibitem{STT-review}
D. C.  Ralph and M. D. Stiles,
J. Magn. Magn. Mater. {\bf 320}, 1190 (2008).

\bibitem{SOT-review}
P. Gambardella and I. M. Miron, Philos. Trans. R. Soc.
London A 369, 3175 (2011).

%\bibitem{Bernevig2005}
%B. A. Bernevig and O. Vafek, Phys. Rev. B {\bf 72}, 033203 (2005).

\bibitem{Pareek2007}
T. P. Pareek, Phys. Rev. B {\bf 75}, 115308 (2007).

\bibitem{Manchon2009}
A. Manchon and S. Zhang,
Phys. Rev. B. {\bf 78}, 212405 (2008); {\bf 79}, 094422 (2009).

\bibitem{Miron2010}
I. M. Miron {\it et al}., Nature Material, {\bf 10}, 230 (2010). 
%G. Gaudin, S. Auffret, B. Rodmacq, A. Schuhl… - Nature materials


%\bibitem{Yokoyama2010}
%T. Yokoyama, J. Zang, and N. Nagaosa, 
%Phys. Rev. B {\bf 81}, 241410 (2010).

%\bibitem{Nomura2010}
%K. Nomura and N. Nagaosa,
%Phys. Rev. B {\bf 82}, 161401(R) (2010)



\bibitem{AHE-review}
N. Nagaosa, J. Sinova, S. Onoda, A. H. MacDonald, and N. P. Ong,
Rev. Mod. Phys. {\bf 82}, 1539 (2010).

\bibitem{Berry-review}
D. Xiao, M.-C. Chang, and Q. Niu,
Rev. Mod. Phys. {\bf 82}, 1959 (2010).

\bibitem{Wan2011a}
X. Wan, A. M. Turner, A. Vishwanath, S. Y. Savrasov, Phys. Rev. B {\bf 83}, 205101 (2011).

\bibitem{Burkov2011}
A. A. Burkov, L. Balents, Phys. Rev. Lett {\bf 107}, 127205 (2011).

\bibitem{Weyl-review}
P. Hosur and X.-L. Qi, 
Comptes Rendus Physique {\bf14}, 857 (2013).

\bibitem{Weyl-review2}
A. M. Turner and A. Vishwanath, 
%A. Beyond band insulators: topology of semi-metals and interacting phases. 
arXiv:1301.0330 (2013).

\bibitem{Dirac-review}
O. Vafek and A. Vishwanath,
Annual Review of Condensed Matter Physics {\bf 5}, 83 (2014).

\bibitem{Kurebayashi2014}
D. Kurebayashi and K. Nomura,
J. Phys. Soc. Jpn. {\bf 83}, 063709 (2014).


\bibitem{Streda1982}
P. Str\v{e}da, J. Phys. C {\bf 15}, L717 (1982).

%\bibitem{Note1}
%In general, the Hall conductivity is decomposed to two parts $\sigma^{I}_{xy}$ and $\sigma^{I\!I}_{xy}$ as discussed in Ref. [\onlinecite{Streda1982}]. $\sigma^{I}_{xy}$ can be associated with states on the Fermi surface. $\sigma^{I\!I}_{xy}$, on the other hand, is the contribution of all states below the Fermi energy and is a thermodynamic equilibrium property of the ferromagnet. In most of cases, $\sigma^{I}_{xy}$ corresponds to the extrinsic contribution, while $\sigma^{I\!I}_{xy}$ the intrinsic contribution. The left-hand side of Eq.~(\ref{Streda-formula}) is $\sigma^{I\!I}_{xy}$. In the following argument, we assume that $\sigma^{I\!I}_{xy}$ is dominant and ignore $\sigma^{I}_{xy}$.

%\bibitem{Note2}
%${\bm T}_{\rm CS}$ differs also from the spin-orbit torque which is generated by the exchange interaction and the spin-orbit interaction, in the sense that.
%${\bm T}_{\rm CS}$ can be introduced in the equilibrium states, and thus differs qualitatively also from spin-orbit torque\cite{Manchon2009} in which the non-equilibrium spin density, induced by a current and spin-orbit coupling, exerts the torque on the local magnetization.




\bibitem{TI-review}
M. Z. Hasan and C. L. Kane, Rev. Mod. Phys. {\bf 82}, 3045 (2010);
X.-L. Qi and S.-C. Zhang, Rev. Mod. Phys. {\bf 83}, 1057 (2011).


\bibitem{Yu2010}
R. Yu, W. Zhang, H.-J. Zhang, S. Zhang, X. Dai, and Z. Fang, Science {\bf 329}, 61 (2010).

%\bibitem{Wolf2001}
%S. A. Wolf et al., Science {\bf 294}, 1488 (2001).

%\bibitem{Jungwirth2006}
%T. Jungwirth et al., Rev. Mod. Phys. {\bf 78}, 809 (2006).


%\bibitem{Wan2011b}
%Wan X, Vishwanath A, Savrasov SY. 2012. Phys. Rev. Lett {\bf 108}:146601

%\bibitem{Xu2011}
%Xu G, Weng H, Wang Z, Dai X, Fang Z. 2011. Phys. Rev. Lett {\bf 107}:186806


%\bibitem{Yu2010}
%R. Yu, W. Zhang, H.-J. Zhang, S. Zhang, X. Dai, and Z. Fang, Science 329, 61 (2010).


\bibitem{ex_BiSe-mag}
Y. L. Chen et al., Science {\bf 329}, 659 (2010)

\bibitem{JAP2012}
X. F. Kou, W. J. Jiang, M. R. Lang, F. X. Xiu, L. He, Y. Wang, Y. Wang, X. X. Yu, A. V. Fedorov, P. Zhang, and K. L. Wang,
J. Appl. Phys. {\bf 112}, 063912 (2012).


\bibitem{J.Zhang2013}
J. Zhang et al., Science {\bf 339}, 1582 (2013).

\bibitem{note3}
Both band gap vanishing and the ferromagnetic order have been observed in [\onlinecite{J.Zhang2013}].

\bibitem{BiSe-model}
H. Zhang, C.-X. Liu, X.-L. Qi, X. Dai, Z. Fang, and S. C. Zhang, Nat. Phys. {\bf 5}, 438 (2009).

\bibitem{BiSe-model2}
C.-X. Liu, X.-L. Qi, H. Zhang, X. Dai, Z. Fang, and S. C. Zhang, Phys. Rev. B {\bf 82}, 041522 (2010). 

\bibitem{Weyl-exp1}
S.-Y. Xu, I. Belopolski, N. Alidoust, M. Neupane,
G. Bian, C. Zhang, R. Sankar, G. Chang, Z. Yuan, C.-C. Lee, S.-M. Huang, H. Zheng, J. Ma, D. S. Sanchez, B. Wang, A. Bansil, F. Chou, P. P. Shibayev, H. Lin, S. Jia, M. Z. Hasan,
Science {\bf 349}, 613 (2015).


\bibitem{Weyl-exp2}
S. Borisenko, D. Evtushinsky, Q. Gibson, A. Yaresko, T. Kim, M. N. Ali, B. Buechner, M. Hoesch, R. J. Cava,
 arXiv:1507.04847 (unpublished)


%\bibitem{Weyl_magnon}
%J. A. Hutasoit, J. Zang, R. Roiban, and C.-X. Liu, arXiv Prepr. arXiv1405.0491 8 (2014).

\bibitem{Yang2011}
K.-Y. Yang, Y.-M. Lu, and Y. Ran, Phys. Rev. B {\bf 84}, 075129 (2011).

\bibitem{Burkov2012}
A. A. Zyuzin and A. A. Burkov, Phys. Rev. B {\bf 85}, 165110 (2012).


\bibitem{Burkov2013}
Y. Chen, S. Wu, and A. A. Burkov, Phys. Rev. B {\bf 88}, 125105 (2013).

\bibitem{Goswami2013}
P. Goswami and S. Tewari, Phys. Rev. B, {\bf 88}, 245107 (2013).



\bibitem{Fujikawa}
K. Fujikawa, Phys. Rev. Lett. {\bf 42}, 1195 (1979).

\bibitem{Burkov2014}
A. A. Burkov, 
Phys. Rev. Lett. 113, 187202 (2014)


\bibitem{Garate2010}
I. Garate and M. Franz,
Phys. Rev. Lett. {\bf 104}, 146802 (2010)

\bibitem{Nomura2010}
K. Nomura and N. Nagaosa,
Phys. Rev. B {\bf 82}, 161401 (2010).
\bibitem{Tserkovnyak2015}
Y. Tserkovnyak, D. A. Pesin, and D. Loss
Phys. Rev. B {\bf 91}, 041121 (2015).
\bibitem{Yokoyama2010}
T. Yokoyama, J. Zang, N. Nagaosa,
Phys. Rev. B {\bf 81}, 24141 (2010).
%\bibitem{Timm2012}
%C. Timm,
%Phys. Rev. B {\rm 86}, 155456 (2012).
\bibitem{Ferreira}
G. J. Ferreira and D. Loss,
Phys. Rev. Lett. {\bf 111}, 106802 (2013).
%\bibitem{Ertler}
%C. E. Ertler, M. Raith, and J. Fabian,
%Phys. Rev. B {\rm 89}, 075432 (2014).
\bibitem{Nature2014}
A. R. Mellnik, J. S. Lee, A. Richardella, J. L. Grab, P. J. Mintun, M. H. Fischer, A. Vaezi, A. Manchon, E.-A. Kim, N. Samarth,  D. C. Ralph,
Nature {\bf 511}, 449 (2014)
\bibitem{Shiomi2014}
Y. Shiomi, K. Nomura, Y. Kajiwara, K. Eto, M. Novak, Kouji Segawa, Yoichi Ando, and E. Saitoh
Phys. Rev. Lett. {\bf 113}, 196601 (2014).
\bibitem{NatureMat2014}
Y. Fan, P. Upadhyaya, X. Kou, M. Lang, S. Takei, Z. Wang, J. Tang, L. He, L.-T. Chang, M. Montazeri, G. Yu, W. Jiang, T. Nie, R. N. Schwartz, Y. Tserkovnyak and K. L. Wang,
Nature Materials {\bf 13} 699 (2014).


%\bibitem{DSM1}
%Z. K. Liu, B. Zhou, Y. Zhang, Z. J. Wang, H. M. Weng, D. Prabhakaran, S.-K. Mo, Z. X. Shen, Z. Fang, X. Dai, Z. Hussain, and Y. L. Chen, Science {\bf 343}, 864 (2014).

%\bibitem{DSM2}
%M. Neupane, S.-Y. Xu, R. Sankar, N. Alidoust, G. Bian, C. Liu, I. Belopolski, T.-R. Chang, H.-T. Jeng, H. Lin, A. Bansil, F. Chou, and M. Z. Hasan, Nat. Commun. {\bf 5}, 3786 (2014).




%\bibitem{CME}
%G. Basar, D. E. Kharzeev, and H.-U. Yee, Phys. Rev. B {\bf 89}, 035142 (2014).



\end{thebibliography}

\end{document}